\documentclass[prb,twocolumn,superscriptaddress,showpacs]{revtex4-1}
\usepackage{epsfig}
\usepackage{dcolumn}
\usepackage{epstopdf}
\usepackage{graphicx}
\usepackage{color}
\begin{document}
\title{Doping-induced dimensional crossover and thermopower burst in Nb-doped SrTiO$_3$ superlattices}
\author{P. Delugas}\author{A. Filippetti}
\affiliation{CNR-IOM, UOS Cagliari, S.P. Monserrato-Sestu Km. 0.700, Monserrato (CA), Italy}
\author{M. J. Verstraete} 
\affiliation{D\'epartement de Physique, B5a, Universit\'e de Li\`ege, B-4000 Sart-Tilman, Belgium}
\author{I. Pallecchi}\author{D. Marr\'e} 
\affiliation{CNR-SPIN UOS Genova and Dipartimento di Fisica, Via Dodecaneso 33, 16146 Genova, Italy}
\author{V. Fiorentini}
\affiliation{CNR-IOM, UOS Cagliari and Dipartimento di Fisica, Universit\`a di Cagliari, Monserrato (CA), Italy}
\date{\today}
\begin{abstract}
Using advanced ab-initio calculations, we describe the formation and confinement of a two-dimensional electron gas in short-period ($\simeq$4 nm) Nb-doped SrTiO$_3$ superlattices as function of Nb doping. We predict complete two-dimensional confinement for doping concentrations higher than 70\%. In agreement with previous observations, we find a large thermopower enhancement at room temperature. However, this effect is primarily determined by dilution of the mobile charge over a multitude of weakly occupied bands. As a general rule, we conclude that thermopower in similar heterostructures will be more enhanced by weak, rathern than tight spatial confinement.

\end{abstract}

\maketitle

\section{Introduction}
 
Since the discovery of 2-dimensional (2D) electron gas (2DEG) in SrTiO$_3$/LaAlO$_3$,\cite{ohtomo} the search of oxide heterostructures with charge-confinement characteristics has been relentlessly pursued by the solid state community. Among the many qualities attributed to 2DEGs, one of the most appealing is the large thermoelectric power. There is mounting evidence, indeed, that nanostructured systems,\cite{venka,harman,majumdar,dress,vining,biswas} rather than bulk materials, can provide a new generation of highly efficient thermoelectric devices capable to directly convert temperature (T) gradients into electric power, and viceversa, thus providing efficient heating and cooling functionalities.\cite{bell}

Recently, large thermopower was observed in several delta-doped SrTiO$_3$ (STO) superlattices.\cite{ohta_nm, jalan} In the 20\% Nb-doped SrTiO$_3$ (STO) superlattices (SLs),\cite{ohta_nm,mune,ohta_tsf} alternating $n$ layers of insulating STO with $m$ layers of 20\% Nb-doped STO (STO$_n$/Nb-STO$_m$) the measured in-plane thermoelectric power, or Seebeck coefficient S, is several times larger than in STO bulk at the same doping. This was hypothesized as due to a density of states (DOS) increase induced by 2D localization.\cite{hicks,sofo,broido,usui} However, this scenario remains to be proved since, in absence of a microscopic description of the system, the presence of a 2D-confined electron gas cannot be assessed. Furthermore, the multi-band nature of transport in oxide heterostructures may give rise to quite a complicated thermoelectric behavior, as seen e.g. for SrTiO$_3$/LaAlO$_3$,\cite{filippetti} whose understanding requires the detailed microscopic description of the heterostructure. 

In this article we describe the 10-layer STO$_9$/Nb-STO$_1$ SL, formed by alternating one Nb-doped layer with a barrier of 9 undoped STO layers at varying Nb-doping concentration. This SL was first considered in the experimental work of Ref.\onlinecite{ohta_nm}, while later works by the same authors\cite{mune,ohta_tsf} extended the study to SLs with a varying number of layers, but always keeping 20\% Nb-doping. Here we study, fully from first principles, three Nb-doping concentrations (25\%, 50\%, and 100\% doping) which are all relevant for experiments since pulsed laser deposition of Nb-doped STO is achievable in the whole 0-100\% doping range.\cite{tomio} Our study is then extended to generic Nb-doping concentration by the use of a multiband effective mass model.

We show that the Nb concentration directly controls the properties of the electron gas. In particular, for large enough nominal doping, a fully confined 2DEG is formed in this short (10-layer period) SL. Furthermore, in agreement with experiment, the Seebeck in the SL is larger than in STO bulk at the same nominal doping. Our space-resolved analysis of thermopower shows that the major increase in thermopower should be attributed to the redistribution of mobile charge in the many bands accessible at finite temperature, i.e. to the charge dilution across a STO region of several nm thickness, so that the increased confinement at high doping ends up being detrimental to thermopower. This agrees with the arguments of Ref.\onlinecite{jalan}, where the  large observed thermopower for La-doped STO SLs was related to the spilling of charge carriers out of the doped region. Our results indicate that, as a general rule, in multiband systems a weak 2D confinement is more conducive to large thermopower than strong 2D confinement. 

\section{Methods: Beyond-LDA band structures combined with Bloch-Boltzmann approach}
\label{secmeth}

To describe the SL we use the ab initio variational self-interaction-corrected density-functional approach (VPSIC),\cite{vpsic} successfully applied to many oxides including STO/LAO \cite{delugas,filippetti} and LNO/LAO \cite{puggioni} superlattices. This approach corrects band gap errors of standard local density functionals, and provides accurate relative band positions and alignments whose inaccuracy would severely compromise predictions for transport in SLs. In particular for what concerns our considered oxide heterostructures, an important quality of VPSIC is accounting accurately for the occupation-dependent band energies related to the confined Ti 3$d$ orbitals. Furthermore, we describe doping in the SL by actual atomic substitutions and explicitly recalculate all properties (including atomic relaxations and electronic structure) at each doping. This is mandatory because the rigid band approximation typically fails in oxide heterostructures. For the bulk, full atomic relaxations are performed at 25\% doping; transport properties at different dopings are obtained using the rigid band approximation, which works well for the bulk.

For the determination of the Seebeck (S) coefficient in diffusive regime, we employ the well known Bloch-Boltzmann transport equations solved in relaxation time approximation (BBT), as implemented in the BoltzTraP code.\cite{madsen} The BBT requires two main ingredients as an input: the electronic band structure and the relaxation time $\tau$. The band structures are calculated by VPSIC on very dense k-space grids (30$\times$30$\times$30 corresponding to 680 k-points in the IBZ for STO bulk, and 20$\times$20$\times$3 giving 230 k-points in the IBZ for the SLs) and interpolated by the linear-tetrahedron approach. The relaxation time $\tau$ tipically depends on carrier energy $\epsilon$ and temperature, and is overwhelmingly difficult to calculate ab-initio for a generic scattering regime, so that it is often assumed to be constant. Within constant relaxation time (CRT) the calculation is quite simplified since $\tau$ cancels out of the expression of Seebeck and Hall resistivity, thus making these two quantities parameter-free and fully determined by the band structure alone. As a further bonus, for constant $\tau$ the Hall factor r$_H$=$\langle\tau^2\rangle/\langle\tau\rangle^2$ (where $\langle\rangle$ indicates average over energy) is equal to unity, and in turn Hall and conduction mobility ($\mu_H$ = $\mu$ r$_H$) become identical, and the Hall resistivity (R$_H$=r$_H$/(n$_{3D}$e)) is simply the inverse of the 3D charge density. 

While very computationally favourable, CRT is rather unsatisfying in terms of quantitative agreement with measurements  (as shown in the next Section). It is therefore necessary to use an energy- and temperature-dependent expression for $\tau$ which could a) overcome the gross disagreement with the experiment, and b) depend on the lowest possible number of parameters, and c) be simple enough to keep calculations feasible even for large-size systems such as oxide heterostructures. Here we adopt for $\tau$ a simple ansatz suggested in the literature,\cite{durc, okuda} based on the factorization in temperature-dependent and energy-dependent parts:
\begin{equation}
{\tau}({\epsilon},T) = F(T) \left( {\epsilon-\epsilon_0 \over K_B T} \right)^{\lambda},
\label{tau}
\end{equation}
where $\epsilon_0$ is the conduction band bottom, $\lambda$ a phenomenological parameter, and F(T) an energy-independent prefactor. The unknown prefactor F(T) cancels out in the expression of Seebeck and Hall resistivity, thus we are left with $\lambda$ as the only parameter. Herafter we will fix $\lambda$=3/2, which optimally reproduces the Seebeck measurement in the whole temperature range (this was previously noticed in Ref.\onlinecite{okuda} where Eq.\ref{tau} is used in combination with an effective-mass model expression of S). Some confusion may result from  the fact that $\epsilon^{3/2}$ is the leading term (for the low-doping regime) of the Brooks-Herring expression of ${\tau}$ for ionized-impurity scattering. The latter mechanism is hardly dominant in STO above 100 K, where polar-optical phonon scattering should be expected. However, Eq.\ref{tau} is radically different from the Brooks-Herring formula, which has a more complicated T-depencence through the Debye screening length and cannot be reduced to the form given in Eq.\ref{tau}. In other words, $\lambda$ in Eq.\ref{tau} should be interpreted as a purely phenomenological fitting parameter, and its effect on the calculated $S(T)$ as unrelated to the predominance of a specific scattering mechanism. In fact, we will show in Section \ref{mbsec} that the main features resulting from our analysis of thermopower are not affected by the specific choice of $\lambda$.
 
Adopting Eq.\ref{tau} the BBT calculation thus remains at the same level of a mere CRT approximation. And yet, it will be shown that use of Eq.\ref{tau} is capable to greatly improve the CRT results for STO-based systems. We expect that a similar improvement could be obtained for wide-gap oxides in general.

\section{STO bulk}

\begin{figure}
\centerline{\includegraphics[clip,width=9.0cm]{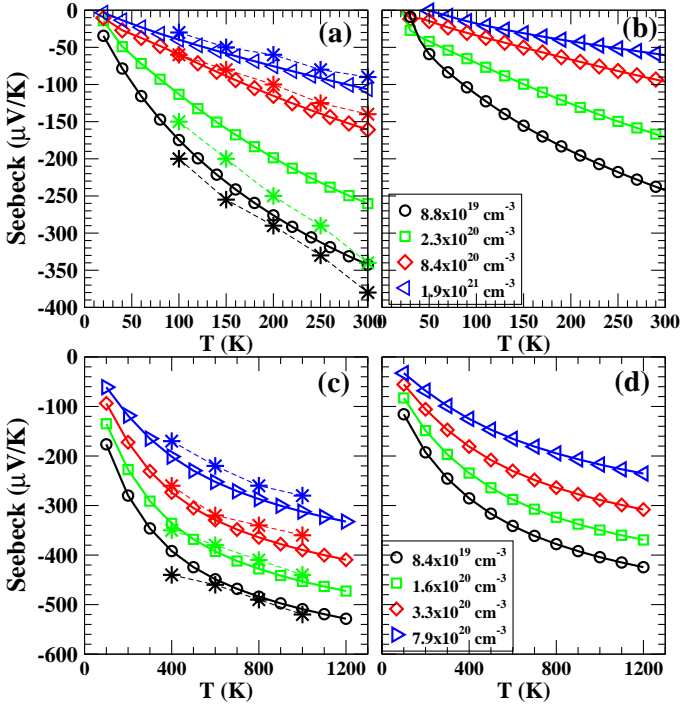}}
\caption{Seebeck calculated by BBT approach for STO bulk. Left panels: calculations for $\tau$ given in Eq.\ref{tau} with $\lambda$=3/2; right panels: calculation with $\lambda$=0 (i.e. constant $\tau$). Top-panels refers to doping concentrations reported in Ref.\onlinecite{okuda}, bottom panels to the concentrations reported in Ref.\onlinecite{s_ohta}. To facilitate the comparison with these experiments, some experimental data (star symbols) extracted 'by-hand' from the figures of the original articles is also included.
\label{seeb}}
\end{figure}

To validate our methodology, we first consider the transport properties of doped bulk STO, that are well known from a number of experiments. For thermopower measurements, we compare our results to two detailed works: Ref.\onlinecite{okuda} for low-T data (below 300 K), and Ref.\onlinecite{s_ohta} for high-T data (up to 1200 K). Our BBT results for S(T) obtained using Eq.\ref{tau} with ${\lambda}$=3/2 and ${\lambda}$=0 (CRT approximation) are shown in Fig.\ref{seeb} for selected doping values matching those reported in the above experimental works (Fig.2 of Ref.\onlinecite{okuda} and Fig.1 of Ref.\onlinecite{s_ohta}).

The comparison clearly demonstrates that our analytic modeling of $\tau$ (Fig.\ref{seeb}, panels (a) and (c)) produces a dramatic improvement over the CRT approximation (Fig.\ref{seeb}, panels (b) and (d)). In the latter, S appears visibly underestimated in absolute value, and its temperature dependence is less structured than the measurements. On the other hand, the adoption of energy-dependent $\tau$ restores a good qualitative agreement with the experiment for a wide range of doping values. Even quantitatively the match with the experiments is rather satisfying, also considering the uncertainty in the actual carrier concentration reported in the experiment (as discussed below in the analysis of Hall resistivity). An exception to this good match is the negative phonon-drag peak at T=50 K measured for the least doped sample of Ref.\onlinecite{okuda} (see Fig.2 of Ref.\onlinecite{okuda}), but this is expected as phonon-drag is not implemented in our BBT calculation, which at present only includes the diffusive term. It is remarkable, nevertheless, that the same value of $\lambda$ can interpolate two sets of measurements obtained in distinct experiments for a very different range of temperatures. This testifies the good transferability of the model, at least for what concerns wide-gap insulating oxides.

\begin{figure}
\centerline{\includegraphics[clip,width=9.0cm]{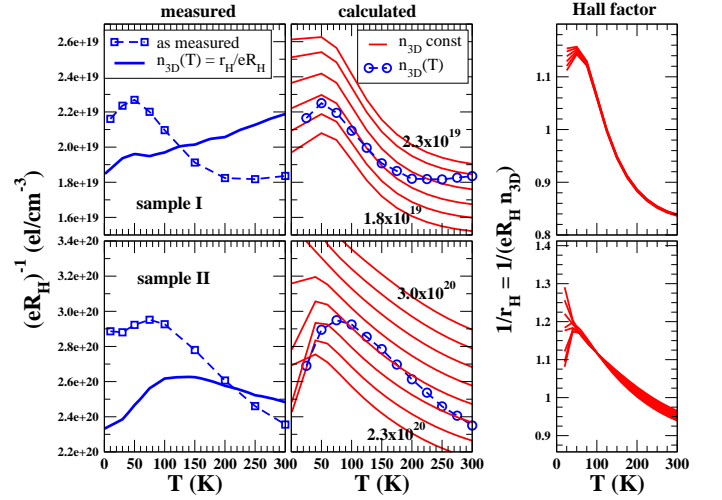}}
\caption{Left panels: measured inverse Hall resistivity $(eR_H)^{-1}$ for two STO bulk samples, a lightly-doped sample I (top) and a heavily-doped sample II (bottom panel).
Blue squared symbols show $(eR_H)^{-1}$ as measured, blue solid line the carrier density n$_{3D}$(T)=$r_H/(eR_H)$ obtained rescaling the measured Hall resistivity by the calculated Hall factor r$_{H}$.  Central panels, red lines: $(eR_H)^{-1}$ calculated for a range of fixed densities spanning the experimental doping range: from 1.8$\times$10$^{19}$ cm$^{-3}$ to 2.3$\times$10$^{19}$ cm$^{-3}$ with incremental steps of 0.1$\times$10$^{19}$ for sample I (top); from 2.3$\times$10$^{20}$ cm$^{-3}$ to 3.0$\times$10$^{20}$ cm$^{-3}$ with increments of 0.1$\times$10$^{20}$ for sample II (bottom). Central panels, blue circlets: $(eR_H)^{-1}$ calculated for the variable charge density n$_{3D}$(T) given by the solid line in the left panel, to be directly compared with the measured vaues (blue squares) to the left. Right panels: Hall factors $r_H$=$(eR_H)n_{3D}$ obtained rescaling $eR_H$ calculated at fixed n$_{3D}$ (red curves of central panels) with these densities. Clearly $r_H$ is weakly dependent on n$_{3D}$, but strongly T-dependent. The average $r_H$ over these densities is used to rescale the measured Hall resistivity and determines n$_{3D}$(T) in the left panels.
\label{hall}}
\end{figure}

A further important quality check of Eq.\ref{tau} is Hall resistivity, which, like S, does not depend on the prefactor F(T) and hence can be calculated plugging just the energy-dependent part of Eq.\ref{tau} into the BBT. Refs.\onlinecite{okuda} and \onlinecite{s_ohta} do not report R$_H$ measurements. We  thus compare calculations with our own Hall measurement for two STO bulk samples (previously used in Ref.\onlinecite{filippetti}) corresponding to two different ranges of doping concentration (the match between calculated and measured S for these samples was already shown to be excellent in Ref.\onlinecite{filippetti}). In Fig.\ref{hall} (left panels) we report $(e\,R_H)^{-1}$ measured for the two samples below T=300 K, to be compared with the calculated values (central panels). The shape of calculated and measured values are nicely similar for both samples, however a direct quantitative comparison is complicated by the dependence of $(e\,R_H)^{-1}$ on the carrier concentration $n_{3D}$, which in the calculation is constant with T and fixed by construction, while in the experiment is unknown and tipically varying with T. To circumvent this ambiguity, we proceeded as following: i) $(e\,R_H)^{-1}$ is calculated (red lines of central panels) for a range of fixed doping values spanning the experimental range of $(e\,R_H)^{-1}$ (for the first sample from 1.8$\times$10$^{19}$ cm$^{-3}$ to 2.3$\times$10$^{19}$ cm$^{-3}$, for the second samples from 2.3$\times$10$^{20}$ cm$^{-3}$ to 3.0$\times$10$^{20}$ cm$^{-3}$). ii) From each of these curves we can easily evaluate the Hall factor as r$_H$ = $(e\,R_H\,n_{3D})$ (red curves in the right panels). According to effective-mass models, we expect r$_H$ to depend only on $\lambda$, and be equal to unity for $\lambda$=0. Indeed, our calculated r$_H$ is almost independent on the density (except at low temperature), and very different from unity, as expected having used $\lambda$=3/2. iii) The r$_H$ average over the considered range of densities is calculated, and then used to rescale the measured $(e\,R_H)^{-1}$ and obtain an estimate of the true carrier concentration as a function of T for the two considered samples (left panels, blue lines). iv) Finally, we can use this estimate of $n_{3D}(T)$ to recalculate $(e\,R_H)^{-1}$ at varying charge density, thus now directly comparable with the experiment (squared symbols in central panel).

We can appreciate the excellent quantitative agreement of calculated and measured Hall resistivity for both samples in the whole temperature range, apart for T lower than 25 K (at low temperature the BBT numerical integration requires extremely dense k-point grids, thus numerical accuracy is very difficult to achieve). We emphasize that it is customary in literature to discard the Hall factor and present the measured $(eR_H)^{-1}$ (with its non-monotonic behaviour as a function of T) as the Hall-measured charge carrier density. Once renormalized by the Hall factor, the estimated carrier density display a more plausible thermally-activated increase with temperature. 
  
In conclusion, our calculation for Seebeck, Hall resistivity, and Hall factors based on Eq.\ref{tau} show a nice quantitative agreement with the experiments and a dramatic improvement over CRT results at null increase of computing cost. This validates the application of the method to the Nb-doped STO SLs, presented in the following.

\section{STO Superlattice}

\subsection{Electronic properties}

\begin{figure}
\centerline{\includegraphics[clip,width=8.0cm]{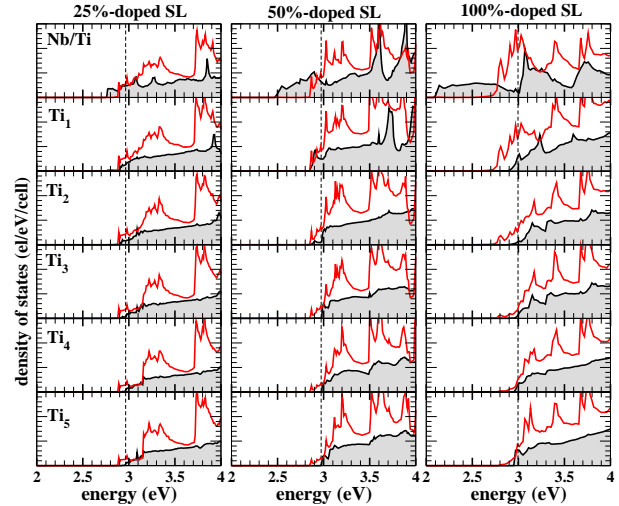}}
\caption{Nb- and Ti-projected  DOS of conduction t$_{2g}$ states in the STO$_9$/Nb-STO$_1$ superlattice at 25\%, 50\%, and 100\% Nb concentration (gray shaded lines: d$_{xy}$; red lines: d$_{xz}$+d$_{yz}$). Top panel is  the doped layer, lowest panel the STO layer  furthest from the doped side. Dashed lines: Fermi energy (energy zero: valence band top).
\label{dos}}
\end{figure}

The DOS of STO$_9$/Nb-STO$_1$ SL at 25\%, 50\%, and 100\% Nb doping is reported in Fig.\ref{dos}. At 100\% doping, the Ti-substituting Nb donates one electron per unit cell area to the SL conduction bands, but the strongly electronegative Nb$^{1+}$ ion keeps most of the mobile charge to itself. As evident from the Figure, at 100\% doping a large portion (0.75 electrons) of this charge remains in the 3d orbitals of the doped layer, 40\% of which in the planar d$_{xy}$ and 30\% in each of the d$_{xz}$ and d$_{yz}$ orbitals, separated from d$_{xy}$ by an energy ${\Delta}t_{2g}$=0.66 eV. While the planar d$_{xy}$ charge is almost completely confined in the doped layer, about half the d$_{xz}$ plus d$_{yz}$ charge (0.25 electrons) spills out into STO as well, as those orbitals propagate along $z$. However this charge fades out rapidly while moving away from the doped plane, and substantially vanishes inside STO. Thus, at large doping our results confirm the presence of a 2DEG confined within a few STO layers, with electronic properties qualitatively similar to those found in STO/LAO.\cite{ohtomo}

We expect that the confinement of d$_{xz}$ and d$_{yz}$ charge will progressively die out as doping decreases, being induced by Nb electronegativity. Indeed, at 50\% doping the d$_{xz}$ and d$_{yz}$ DOS are almost evenly distributed through STO, although with some remnant accumulation near the doped layer. The d$_{xy}$ charge, on the other hand, still fully belongs to the 50\%-doped layer. At 25\% doping (close to experimental 20\%) the d$_{xz}$ and d$_{yz}$ charge is homogeneously spread throughout the SL with no residual accumulation near Nb layer, while the d$_{xy}$ charge is still 2D.

\begin{figure}
\centerline{\includegraphics[clip,width=8.0cm]{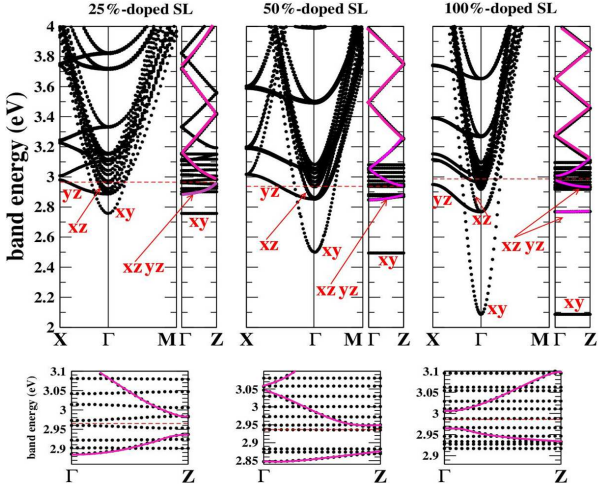}}
\caption{Top: bands of the STO$_9$/Nb-STO$_1$ SL at 25\%, 50\%, and 100\% doping. Dashed lines are Fermi energies; energy zero is placed at the valence band top. The character of the three lowest bands is labeled. The conduction bands of d$_{xz}$, d$_{yx}$ character along $\Gamma$-Z=[001] are highlighted in violet: with increasing Nb doping, a gap opens between the flat lowest branch and the higher downfolded bulk-like sections. Bottom: enlargement of the bands around $\epsilon_F$. 
\label{band}}
\end{figure}

The doping-controlled dimensional crossover involving the three lowest bands of the SL is even more explicit in the band structure (Fig.\ref{band}): at low doping the d$_{xz}$ and d$_{yz}$ bands are bulk-like, but as doping increases the lowest one progressively flattens out, with a gap opening to the higher bulk-like bands. The effective mass of the lowest band $m_{xz,[001]}^*$=$m_{yz,[001]}^*$ increases from 0.39 to 0.83 to 3.85 (in electron mass units) for 25\%, 50\%, and 100\% doping (the corresponding mass is 0.32 in bulk SrTi$_{0.75}$Nb$_{0.25}$O$_3$). On the other hand, the lowest d$_{xy}$ band is fully confined at any doping, with $m_{xy,[001]}^*$$\simeq$1000 compared to 5.45 in SrTi$_{0.75}$Nb$_{0.25}$O$_3$. A zoom near the Fermi energy ($\epsilon_F$) (Fig.\ref{band} bottom) shows that the SL spectrum is actually gapped along $k_{z}$; a non-vanishing conductivity at room temperature is still expected, however, because of the high DOS near $\epsilon_F$.

Our results thus far describe this SL as a double-channel conduction system, with a portion of charge of d$_{xy}$ orbital character fully confined in 2D at any doping concentration, and a fraction of d$_{xz}$, d$_{yz}$ charge which may be 2D or 3D in nature depending on the doping concentration. As described in the following, these two channels will contribute differently to in-plane thermopower.

\subsection{Thermopower}

\begin{figure}
\centerline{\includegraphics[clip,width=7.5cm]{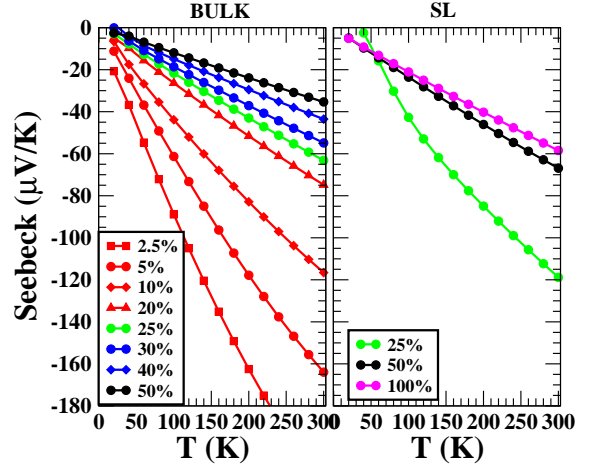}}
\caption{Bloch-Boltzmann-calculated planar Seebeck S vs T for STO bulk (left) and the STO$_9$/Nb-STO$_1$ SL (right panel) at various Nb-doping concentrations. For bulk, calculations are actually done at 25\% doping, where data at other dopings are calculated in a rigid-band approach, which is acceptable in the bulk.
\label{seeb_vs_t}}
\end{figure}

We use the calculated band energies as input for the Bloch-Boltzmann transport theory,\cite{madsen} and calculate (Fig.\ref{seeb_vs_t}) the in-plane components of Seebeck coefficient (S) as a function of temperature for the Nb-doped STO bulk and the 10-layer SL at varying Nb doping concentration.  At T=300 K, the calculated Seebeck for the SL (S$_{\rm{sl}}$) is enhanced by about a factor 2 over that of the STO bulk (S$_{\rm{bulk}}$) at same nominal doping, in qualitative agreement with the experiment.\cite{ohta_nm,mune} Specifically, our $|$S$_{\rm{bulk}}|$=60 $\mu$V/K at 25\% doping is close to 62 $\mu$V/K measured \cite{mune} at 20\%; however, our $|$S$_{\rm{sl}}|$=120 $\mu$V/K  at 25\% is half the experimental\cite{mune} 240 $\mu$V/K at 20\% doping. The discrepancy may be due to defects or stoichiometry fluctuations which may reduce, with respect to nominal doping, the effective mobile charge contributing to transport, similarly to what happens in STO/LAO.\cite{ohtomo} Indeed, our S$_{\rm{sl}}$ at low doping (see Fig.\ref{model} below) matches the experimental value at $\sim$8\% Nb doping, corresponding to a density 1.3$\times$10$^{20}$cm$^{-3}$, which is indeed not too far from value 2.2$\times$10$^{20}$cm$^{-3}$ reported in Ref.\onlinecite{mune}.

\begin{figure}
\centerline{\includegraphics[clip,width=6.0cm]{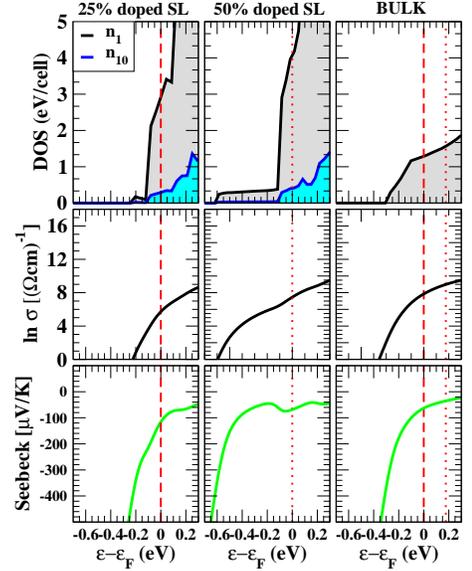}}
\caption{Top: total DOS for the 25\% and 50\% Nb-doped STO$_9$/Nb-STO$_1$ SL and for the 25\%-doped bulk. For the former, two different DOS are shown: n$_{1}$, normalized on a one-layer volume, and  n$_{10}$ normalized on the whole SL volume. Center: planar logarithmic conductivity ($\sigma$) at T=300 K. Bottom: planar Seebeck at T=300 K for the same systems. Red lines indicate Fermi energies for 25\% doping (dashed) and 50\% doping (dotted). 
\label{seeb_300k}}
\end{figure}

We now investigate the reason for the thermopower enhancement. In Fig.\ref{seeb_300k} we show the calculated DOS (n$(\epsilon$), upper panels), the in-plane logarithmic electrical conductivity $\ln\,$$\sigma(\epsilon)$ determined to within an additive term $\ln\,$F(T) (middle panels), and Seebeck (lower panels) as a function of chemical potential at T=300 K for the SL at 25\% and 50\% doping, and for bulk SrTi$_{0.75}$Nb$_{0.25}$O$_3$. 

These results can be analyzed with the help of the Cutler-Mott formula \cite{cutler}
\begin{equation}
{\rm S} = 
\frac{\pi^2k_B^2T}{3e}\frac{\partial (\ln\,{\sigma})}{\partial\epsilon}|_{\epsilon_F} = 
%\frac{\pi^2k_B^2T}{3e}\frac{1}{n}\frac{\partial n}{\partial\epsilon}|_{\epsilon_F},
%\frac{\pi^2k_B^2T}{3e}\left[\frac{\partial (\ln\,n)}{\partial\epsilon}|_{\epsilon_F}+\frac{\partial (\ln\,\mu)}{\partial\epsilon}|_{\epsilon_F}\right]
\frac{\pi^2k_B^2T}{3e}\left|\frac{\partial (\ln\,n)}{\partial\epsilon}+\frac{\partial (\ln\,\mu)}{\partial\epsilon}\right|_{\epsilon=\epsilon_F}
\label{cutlermott}
\end{equation}
where $\sigma(\epsilon)$=$e n(\epsilon)\mu(\epsilon)K_BT$, and $\mu(\epsilon)$ are spectral conductivity\cite{note} and mobility, respectively.

Our BBT results in Fig.\ref{seeb_300k} are quite consistent with Eq.\ref{cutlermott}, see values in Table \ref{tab}: The logarithmic derivatives of spectral conductivity in the SL are about twice that in bulk, and hence so is S. Eq.\ref{cutlermott} helps further in explaining the difference between S$_{\rm{bulk}}$ and S$_{\rm{sl}}$. If the SL charge were entirely confined in the doped layer, the relevant DOS entering the expression for S would be $n_1$ (see Fig.\ref{seeb_300k}, top) normalized to the volume of a single layer. The slope of $n_{1}$ increases markedly compared to the bulk, indicating  a genuine increase of charge localization. However, we have previously demonstrated that the charge spreads through the whole SL at any doping, thus the density $n_{10}$ normalized to the whole 10-layer SL volume is the correct choice for the SL. At $\epsilon_F$ the slopes of $n_{10}$ and of the bulk DOS at same Nb doping are similar, whereas $n_{10}$ is definitely smaller than the bulk DOS. If we discard the mobility dependence on the energy in Eq.\ref{cutlermott}, which is typically smaller than the charge density dependence, we conclude that the increase in $(1/n)(\partial n/\partial\epsilon$)$|_{\epsilon_F}$ (hence in Seebeck) should be due to a DOS decrease (i.e. charge dilution through the SL) rather than to a DOS slope increase (i.e. mass enhancement). 

\begin{table}
\begin{center}
\caption{Logarithmic derivative of density of states n$(\epsilon$) and spectral conductivity $\sigma(\epsilon)$ at $\epsilon_F$, and Seebeck S calculated by BBT approach at T=300 K for 25\% and 50\% Nb-doped STO superlattice and STO bulk. 
\label{tab}}
\begin{tabular}{lcccc} 
\hline
		        & SL 25\%  &  SL 50\%  & bulk 25\% & bulk 50\% \\
\hline
\hline
$\partial (\ln\,n)/\partial\epsilon|_{\epsilon_F}$ (eV$^{-1}$)	          & 3.2  & 2.9   & 1.0	 & 1.2 \\		
$\partial (\ln\,\sigma)/\partial\epsilon|_{\epsilon_F}$ (eV$^{-1}$)       & 15.7 & 9.3   & 8.6   & 4.8  \\
S ($\mu$V/K)                                                              & -120 & -68   & -60   & -35  \\
\hline
\end{tabular}
\end{center}
\end{table}

\subsection{Multiband modeling}
\label{mbsec}

\begin{figure}
\centerline{\includegraphics[clip,width=8.5cm]{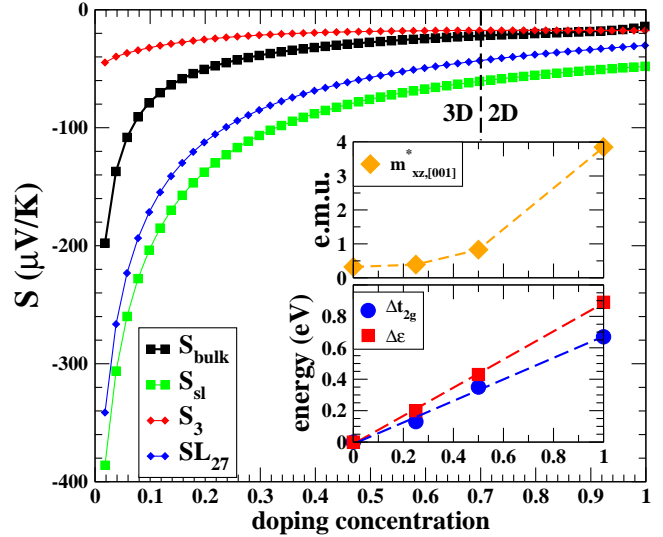}}
\caption{Left: Thermopower as a function of Nb-doping concentration calculated using the multiband effective-mass model with doping-dependent band parameters, for STO bulk (black) and the STO$_9$/Nb-STO$_1$ SL (green line) at T=300 K. For the latter, contributions from the 3 lowest t$_{2g}$ bands (red) and the remaining 27 t$_{2g}$ bands (blue line) are also shown. The vertical dashed line separates regions of 3D (low-doping) and 2D (high-doping) carrier regime (see text). Inset: model parameters interpolation as a function of doping (see text).
\label{model}}
\end{figure}

Direct ab-initio calculations for generic doping values requires a workload beyond current computational capabilities. To further buttress our previous conclusions and generalize our analysis to doping levels not accessible by direct first-principles calculations, we have therefore used a 3-dimensional effective mass modeling (a similar implementation was previously used for STO/LAO\cite{filippetti}) including all the t$_{2g}$ conduction bands of the full calculation (30 bands for the 10-layer SL). In order to include the important changes of the band structure with the doping concentration, this model uses a doping-dependent interpolation of the ab initio VPSIC values for three key quantities (see Fig.\ref{model}, inset): the t$_{2g}$ energy splitting (${\Delta}t_{2g}$) between purely planar d$_{xy}$ and orthogonal d$_{xz}$, d$_{yz}$ states, the energy difference between the lowest d$_{xy}$ band and the bulk-like STO conduction band manifold ($\Delta\epsilon$), and the effective mass of the bands involved in the dimensional crossover (m$_{xz,[001]}^*$=m$_{yz,[001]}^*$). This procedure effectively circumvents the rigid band approximation, avoiding its inaccuracies. The model is validated by its reproducing the Seebeck coefficient obtained directly by Bloch-Boltzmann calculations at 25\%, 50\% and 100\% doping.

In Fig.\ref{model} (main panel) we compare S$_{\rm{bulk}}$ and S$_{\rm{sl}}$ at T=300 K vs Nb concentration. S$_{\rm{sl}}$ is further broken down into contributions from the three lowest bands (S$_{\rm{3}}$) and all the other 27 t$_{2g}$ bands (S$_{\rm{27}}$) included in the model:
\begin{equation}
S_{\rm{sl}}\,=\,S_{\rm{3}}+S_{\rm{27}}\,=\sum_{i=1,3} {\sigma_i S_i\over \sigma}+\sum_{i=4,30} {\sigma_i S_i\over \sigma}
\end{equation}
where $\sigma_i$ and S$_i$ are conductivity and thermopower of the $i$$^{th}$ band. As doping decreases, we see a progressive increase in S$_{\rm{sl}}$ over S$_{\rm{bulk}}$, which follows almost entirely from the enhanced $|$S$_{\rm{27}}|$ contribution. This is easily understood recalling that $|S|$ is inversely related to $\epsilon_F$: at low doping the SL charge can be progressively diluted through a large number of bands, in turn lowering $\epsilon_F$ with respect to the bulk. At zero doping $\Delta\epsilon$ $\sim$0, ${\Delta}t_{2g}$$\sim$0, and the full dilution limit S$_{\rm{27}}$=0.9 S$_{\rm{sl}}$ is reached. On the other hand, $|$S$_{\rm{3}}|$ is always smaller than $|$S$_{\rm{27}}|$ and changes barely with doping, despite the fact that only the two lowest d$_{xz}$, d$_{yz}$ bands are affected by confinement. Indeed, while the 2D confinement (i.e. the increase of m$_{xz,[001]}^*$) in itself lowers $\epsilon_F$, the increment of doping stabilizes the three lowest bands (i.e. enhances $\Delta\epsilon$), thus causing a flow of additional charge from the higher-energy bands and effectively rising $\epsilon_F$; the net effect is that S$_{\rm{3}}$ remains nearly constant with doping, and progressively approaches S$_{\rm{bulk}}$ as Nb doping increases. Above 70\% doping, S$_{\rm{3}}$$\sim$S$_{\rm{bulk}}$  because the charge collapses into the three lowest bands (at T=0), which are now well separated from the undoped STO band manifold. A doping of 70\% is thus the estimated threshold between 3D and 2D behavior. Nevertheless, the thermal occupancy of the higher bands at T=300 K is sufficient to furnish a sizeable S$_{\rm{27}}$ contribution to the total S$_{\rm{sl}}$, still visibly larger than S$_{\rm{bulk}}$.

These results thus indicates that the increase of S$_{\rm{sl}}$ relative to S$_{\rm{bulk}}$ originates from charge dilution through the SL, and not from confinement-induced charge localization. This has a simple rationale: for a single-band system, enhancing the effective mass is tantamount to reducing $\epsilon_F$, in turn increasing the thermopower; but for a multi-band system a very tight 2D confinement may actually cause $\epsilon_F$ to rise, and be detrimental for thermopower compared to a milder confinement allowing 2DEG dilution over a larger thickness.

We underline that charge dilution in confined systems (where mobile charge is inhomogeneously distributed in space) is different from a trivial decrease of carrier density. This can be seen in a very simple case: Suppose the charge n$_{3D}$ initially localized in a single band of d$_{xy}$ character (thus fully confined in a single layer) filled up to $\epsilon_F$, and let us redistribute it into N identical bands filled up to $\epsilon_F'$, all with same mobility $\mu$, charge n$_i$ = n$_{3D}$/N, conductivity $\sigma_i$=$e\,n_i\,\mu$, and Seebeck S$_i$=S($\epsilon_F'$). The  conductivity of the diluted system is of course unchanged: $\sigma$ =$\sum_i \sigma_i$ = N$\sigma_i$, wherease the Seebeck:
\begin{equation}
S=\sum_{i=1,N} {\sigma_i S_i\over \sigma}= S_i(\epsilon_F'),
\end{equation}
must instead be larger than S($\epsilon_F$) since $\epsilon_F'$ is lower than $\epsilon_F$. That is, pure charge dilution in a multitude of degenerate bands always increases the Seebeck and leave conductivity unchanged. These are favourable conditions for good thermoelectric efficiency. Of course, other factors not included in this simple hypothesis may affect this balance, such as changes of effective masses due to genuine charge localization, or changes in the scattering mechanism (hence in $\tau$ and $\mu$). However, it clearly holds as a general guideline that weak 2D confinement is a more favorable condition than tight 2D confinement to obtain large Seebeck values.

\begin{figure}
\centerline{\includegraphics[clip,width=9.0cm]{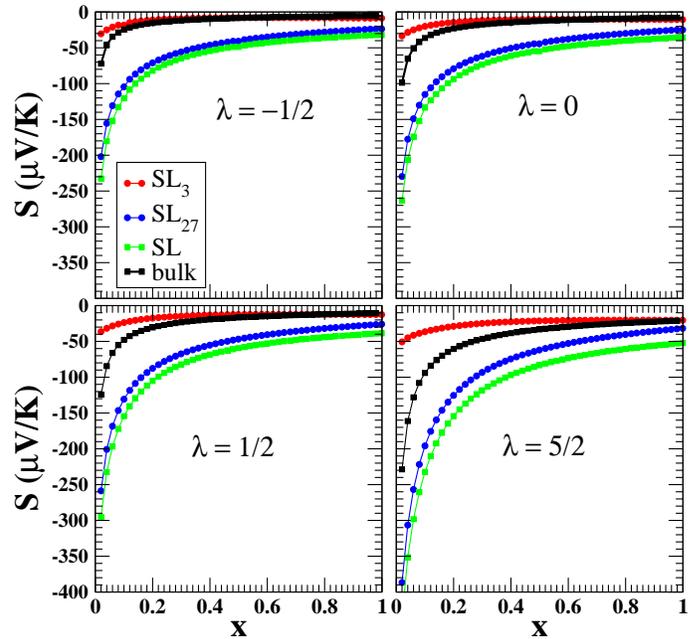}}
\caption{Left: Thermopower as a function of Nb-doping concentration calculated using the multiband effective-mass model with doping-dependent band parameters, for STO bulk (black) and the STO$_9$/Nb-STO$_1$ SL (green line) at T=300 K. For the latter, contributions from the 3 lowest t$_{2g}$ bands (red) and the remaining 27 t$_{2g}$ bands (blue line) are also shown. Results of different panels only differ for the value of the power parameter $\lambda$ in Eq.\ref{tau}.
\label{model2}}
\end{figure}

Finally, in Fig.\ref{model2} we replicate the result for different values of $\lambda$, to give evidence that the fundamental conclusion of this analysis is unaffected by the choice of this parameter. We clearly see that while absolute values of total and band-decomposed Seebeck do depend on $\lambda$, the contribution of the 27 minority-occupied bands is always dominating over the 3 bands of the doped layer. Thus, we can conclude by saying that independently on the scattering regime, charge dilution is always effective in producing an important burst in thermopower.

\section{Conclusions}

In conclusion, we have described from a theoretical viewpoint the characteristics of the electron gas present in short-period $\delta$-doped oxide superlattices. We have shown that the electronic properties (effective mass and spatial extension) of the mobile charge in the SL can be effectively tuned by the diagnostic choice of the doping concentration: above the estimated threshold of 70\% doping, a dimensional crossover takes place, and a fully confined 2DEG appears. Below this threshold, electron charge accumulates near the doped layer, but a consistent fraction of it (progressively increasing with the lowering of doping concentration) spreads through the whole SL, so that a complete 2D confinement is not achieved. We remark that very high Nb-doping concentrations in STO are experimentally achievable, and apparently keen to the reach of high electron mobility.\cite{tomio}

In agreement with experiments,\cite{ohta_nm,jalan} we find the thermopower of the SL remarkably larger than the thermopower of the bulk at equivalent doping concentration. Such an increase of thermopower is found to be consequence of the delocalization of carriers into a multitude of barely occupied bands. This conclusion can be understood considering that, according to the Boltzmann theory, the dominant factor in expanding the Seebeck amplitude is primarily the lowering of the Fermi energy which obviously follows from the dilution.     

As a general rule, our analysis shows that in a charge-confined (thus inhomogeneous) multi-band system, a weak 2D confinement favors large thermopower more than a strong confinement which tightly traps all the charge in one or a few doped layers.

\acknowledgments
Work supported in part by projects EU FP7 {\it OxIDes} (grant n.228989), MIUR-PRIN 2010 {\it Oxide}, IIT-Seed NEWDFESCM, IIT-SEED POLYPHEMO and "platform computation" of IIT, 5 MiSE-CNR, and Fondazione Banco di Sardegna grants. MJV acknowledges a visiting professor grant at Cagliari University, the Belgian ARC project TheMoTherm, and a ``Cr\'edit d'impulsion'' grant from University of Li\`ege. Calculations performed at CASPUR Rome and Cybersar Cagliari.

%%%%%%%%%%%%%%%%%%%%%%%  biblio

%%%%%%%%%%%%%%%%%%%%%%%%%%%%%%%%%%%%%%%%%%%%%%%%%%%%%%%%%%%%%%%%%%%%%%%%%%%%%%%%%%%%


\begin{thebibliography}{99}
\bibitem{note}
From $\sigma{\epsilon}$ the electric conductivity is obtained by integrating over the energy $\sigma(\epsilon)$ times the energy derivative of the Fermi function.
\end{thebibliography}
\end{document}